\documentclass[10pt,english]{article}
\usepackage{graphicx}
\usepackage[T1]{fontenc}
\usepackage[latin9]{inputenc}
\usepackage[margin=1.0in]{geometry}
\usepackage{float}
\usepackage{amsmath}
\usepackage{amsbsy}
\usepackage{setspace}
\usepackage{amssymb}
\usepackage{esint}
\usepackage{cite}
\newfloat{algorithm}{tbp}{loa}
\floatname{algorithm}{Algorithm}
\usepackage{algorithmic}

\newlength\myindent
\newcommand\bindent{
  \begingroup
  \setlength{\itemindent}{\myindent}
  \addtolength{\algorithmicindent}{\myindent}
}
\newcommand\eindent{\endgroup}

\makeatletter

\floatstyle{ruled}
\newfloat{algorithm}{tbp}{loa}
\floatname{algorithm}{Algorithm}

\usepackage{babel}

\begin{document}

\title{Systemic Risk, Maximum Entropy and Interbank Contagion}

\author{M. Andrecut}

\date{April 3, 2016}

\maketitle
{

\centering Calgary, Alberta, T3G 5Y8, Canada

\centering mircea.andrecut@gmail.com

} 
\bigskip 
\begin{abstract}
We discuss the systemic risk implied by the interbank exposures reconstructed with the 
maximum entropy method. The maximum entropy method severely underestimates the risk of 
interbank contagion by assuming a fully connected network, while in reality the structure 
of the interbank network is sparsely connected. Here, we formulate an algorithm for 
sparse network reconstruction, and we show numerically that it provides a more reliable estimation of 
the systemic risk. 


\end{abstract}
\bigskip 

\section{Introduction}

The analysis of interbank contagion has been receiving attention recently due to the increasing deterioration 
of the stability in the interbank lending market \cite{key-1}-\cite{key-10}.
This instability may lead to a domino effect, 
where the failure of a bank may trigger a cascading of failures of other banks, even if they 
are not directly exposed to the initially failing bank. Therefore, it is essential to understand 
the potential contagion mechanisms in order to minimize the systemic risk imposed by an unperforming 
interbank lending network. 

The correct estimation of the risk of contagion suffers from the incomplete knowledge of the 
details regarding the interbank bilateral exposures, which generally are not available because 
the banks do not disclose their bilateral exposures to central banks and regulators. In general,  
only the total interbank assets and liabilities of each bank can be estimated from their balance 
sheet \cite{key-1}-\cite{key-10}. 
Therefore the bilateral exposures, which are essential for risk models, cannot be estimated 
without imposing further assumptions. The standard approach is to estimate the bilateral exposures 
using the Maximum Entropy (ME) method which spreads the exposures as evenly as possible, such 
that it satisfies the constraints corresponding to the total assets and liabilities for each bank \cite{key-1}-\cite{key-10}.
Unfortunately, this method is known to provide an unrealistic network topology, because it assumes 
a fully connected network, while in reality the interbank network is sparsely connected \cite{key-9,key-11,key-12}.
The real sparse network structure is dictated by the fact that banks cannot spread their network linkages across 
the entire system, because maintaining such a large number of connection is obviously costly, and therefore often 
the network is very sparse, with only a small number of established connections \cite{key-12}. As a direct consequence 
of this discrepancy, the ME method severely underestimates the risk of interbank contagion. 

Several algorithms for sparse network reconstruction have been proposed in order to overcome the 
limitations of the ME method \cite{key-11,key-12}. These algorithms are based on heuristic methods and they have a relatively high cost computational requirements. 
Here we formulate another algorithm for sparse network reconstruction, with a much simpler architecture and 
very fast implementation. We show numerically that this algorithm achieves a high degree of network sparsity, 
and using numerical stress-test simulations we show that it provides a more reliable 
risk estimation. 

The remainder of the paper is organized as follows. The standard dense network reconstruction approach is described in Section 2. 
In Section 3 we discuss the sparse reconstruction algorithm and its implementation. The performance of the proposed algorithm is analyzed in Section 4. 
In Section 5 we formulate and simulate the contagion stress-test for the proposed algorithm. The final Section 6 summarizes the main results and the conclusions.

\section{Dense network reconstruction}

We consider a network of $N$ banks $B=\lbrace b_0,\hdots,b_{N-1} \rbrace$, where each bank may borrow 
to or lend money from other banks in $B$. The interbank relationship can be represented by an $N \times N$ 
matrix $x=\left[ x_{ij} \right]_{N \times N} $ where $x_{ij}\geq0$ denotes outstanding loans and deposits of bank $b_i$ 
to bank $b_j$. The sum of the matrix elements across the row $i$ gives the total value of bank $b_i$ assets,
and the sum across the column $j$ gives bank $b_j$ total liabilities as follows:
\begin{equation}
a_i = \sum_{j=0}^{N-1} x_{ij}, \quad \ell_j = \sum_{i=0}^{N-1} x_{ij}.
\end{equation}
Also, without restricting generality we consider a closed economy, such that total interbank assets and liabilities 
are equal:
\begin{equation}
\sum_{i=0}^{N-1} a_{i} = \sum_{j=0}^{N-1} \ell_{j} = \Lambda,
\end{equation}
and therefore the exposures reflect the relative importance of each bank in the interbank network.
Without loss of generality we also assume that $\Lambda = 1$, unless stated otherwise.

The $x$ matrix provides information about the interbank exposures, and in principle it should 
be sufficient to estimate the risk of contagion. However, as mentioned in the introduction the bilateral 
exposures $x_{ij}$ are generally unknown, and only the total assets $a_i$ and liabilities $\ell_j$ are typically 
observable from the balance sheet of each bank. Therefore, the main problem is to estimate the interbank exposures matrix $x$, 
given only the assets $a_i$ and liabilities $\ell_j$, $i,j=0,\hdots,N-1$.

The ME method solves the following optimization problem, subject to the constraints (1):
\begin{equation}
\max_x S_x,
\end{equation}
where 
\begin{equation}
S_x = -\sum_{i=0}^{N-1}\sum_{j=0}^{N-1} x_{ij}\ln x_{ij}.
\end{equation}
is the entropy of the matrix $x$. Using the the method of Lagrange multipliers one can easily show that the solution for this problem is:
\begin{equation}
x_{ij} = a_i \ell_j,\quad i,j=0,\hdots,N-1.
\end{equation}
Intuitively, this solution spreads the exposures as evenly as possible, consistent with the constraints, filling completely the matrix $x$, which is not in agreement  
with the real interbank networks. 

A first improvement is to consider that a bank cannot have an exposure to itself, which means that the diagonal elements of the matrix $x$ 
must be zero:
\begin{equation}
x_{ij}^0 = 
\begin{cases}
a_i \ell_j & i\neq j \\
0 & i = j
\end{cases}.
\end{equation}
Obviously, the matrix $x^0$ can no longer satisfy the imposed constraints (1). However, one can find a solution $x$ that minimizes 
the Kullback-Leibler divergence (also known as the cross entropy or the relative entropy) between $x$ and $x^0$:
\begin{equation}
D(x\parallel x^0) = \sum_{i=0}^{N-1}\sum_{j=0}^{N-1} x_{ij}\ln \frac{x_{ij}}{x_{ij}^0},
\end{equation}
which means that the solution $x$ will be as close as possible to $x^0$. Since $D$ is not defined for $x_{ii}^0=0$, we should note that for $x,\varepsilon>0$ we have:
\begin{align}
\lim_{x \rightarrow 0} x \ln x &= 0, \\
\lim_{\varepsilon \rightarrow 0} x \ln \frac{x}{\varepsilon} &= \infty.
\end{align}

Therefore, the optimization problem subject to the constraints (1) becomes:
\begin{equation}
\min_x D(x\parallel x^0).
\end{equation}
This problem can no longer be solved analytically, and therefore requires numerical optimization. 
The RAS algorithm provides a computationally efficient method to solve this minimization problem \cite{key-13}. 

The algorithm starts by allocating an array $x$ with $N^2$ number of elements, and 
setting $x_{ij}(0) = x_{ij}^0$. The algorithm iterates the following equations:
\begin{align}
x_{ij}(t+1) &= \frac{x_{ij}(t)a_i}{\sum_{n=0}^{N-1} x_{in}(t)},\quad i,j=1,\hdots,N-1\\
x_{ij}(t+1) &= \frac{x_{ij}(t+1)\ell_j}{\sum_{n=0}^{N-1} x_{nj}(t+1)},\quad j,i=1,\hdots,N-1, 
\end{align}
such that a complete iteration consists of two loops, corresponding to the rows and respectively the columns of $x$. 
The algorithm stops when the Euclidean distance $\eta$ between two complete 
iterations is smaller that a prescribed error $0< \delta \ll 1$:
\begin{equation}
\eta = \Vert x(t+1) - x(t) \Vert < \delta.
\end{equation}

The solution matrix $x$ is called the \textit{Maximum Entropy (ME) solution} in the economic literature, 
since it is the closest matrix to the ME matrix (12), and consistent with the imposed constraints (1) \cite{key-1}-\cite{key-12}.
This solution is still dense because only the diagonal of the matrix $x$ is zero, which also 
leads to an unrealistic interbank network structure.

\section{Sparse network reconstruction}

Let us assume that the interbank network is sparse, and it is described by the adjacency matrix $q=\left[ q_{ij} \right] _{N \times N}$.
The adjacency matrix has binary coefficients $q_{ij}\in \lbrace 0,1 \rbrace$, such that $q_{ij}=1$ if there is a relationship 
between the banks $b_i$ and $b_j$, and $q_{ij}=0$ otherwise. 
The connectivity $\kappa \in [0,1]$ and the sparsity $\sigma \in [0,1]$ of the interbank network are therefore given by:
\begin{equation}
\kappa = N^{-2} \sum_{i=0}^{N-1}\sum_{j=0}^{N-1} q_{ij}, \quad \sigma = 1 - \kappa.
\end{equation}
Our goal is to find a matrix $x$ consistent with the constraints (1), such that:
\begin{equation}
q = \Theta (x),
\end{equation}
where $\Theta(x)$ is the Heaviside function applied element-wise:
\begin{equation}
q_{ij} = \Theta (x_{ij})=
\begin{cases}
		1 & \text{if }\; x_{ij}>0 \\
		0 & \text{otherwise}
\end{cases}.	
\end{equation}

We solve this problem by minimizing the Kullback-Leibler divergence between $x$ and $q$:
\begin{equation}
\min_x D(x\parallel q).
\end{equation}
First we define a new set of variables as follows:
\begin{equation}
y_{ij} =
	\begin{cases}
		x_{ij} & \text{if }\; q_{ij} = 1 \\
		0 & \text{if }\; q_{ij} = 0
	\end{cases}
\;\Leftrightarrow\; x_{ij} = q_{ij} y_{ij},
\end{equation}
such that the Lagrangian of the problem becomes:
\begin{align}
L(y_{ij},\alpha_i,\lambda_j) &= \sum_{i=0}^{N-1}\sum_{j=0}^{N-1} q_{ij}y_{ij}\ln y_{ij} \nonumber \\
&+ \sum_{i=0}^{N-1} \alpha_i \left( a_i - \sum_{j=0}^{N-1} q_{ij}y_{ij} \right) \nonumber \\
&+ \sum_{j=0}^{N-1} \lambda_j \left( \ell_j - \sum_{i=0}^{N-1} q_{ij}y_{ij} \right), 
\end{align}
The optimality conditions:
\begin{equation}
\frac{\partial L}{\partial y_{ij}}=0,\quad  \frac{\partial L}{\partial \alpha_i}=0,\quad  \frac{\partial L}{\partial \lambda_j} = 0, \,
\end{equation}
give the following equations:
\begin{equation}
\ln y_{ij} + 1 - \alpha_i - \lambda_j = 0,
\end{equation}
\begin{equation}
\sum_{j=0}^{N-1} q_{ij}y_{ij} = a_i,\quad \sum_{i=0}^{N-1} q_{ij}y_{ij} = \ell_j. 
\end{equation}
From the first equation we have:
\begin{equation}
y_{ij} = \exp \left( \alpha_i + \lambda_j - 1 \right). 
\end{equation}
Here we define the new variables:
\begin{align}
\psi_i &=  \exp \left( \alpha_i - 1/2 \right), \\
\varphi_j &=  \exp \left( \lambda_j - 1/2 \right),
\end{align}
such that:
\begin{equation}
y_{ij} = \psi_i \varphi_j,\quad i,j=0,\hdots,N-1.
\end{equation}
From the constraints we also obtain the following equations:
\begin{align}
\psi_i &= \frac{a_i}{\sum_{j=0}^{N-1} q_{ij}\varphi_j},\quad i=0,\hdots,N-1,\\
\varphi_j &= \frac{\ell_j}{\sum_{i=0}^{N-1} q_{ij}\psi_i},\quad j=0,\hdots,N-1,
\end{align}
which are the core of the sparse reconstruction algorithm. 

The variables $\psi_i$ and $\varphi_j$ can be obtained iteratively from the above equations, as follows:
\begin{align}
\psi_i (t+1) &= \frac{a_i}{\sum_{j=0}^{N-1} q_{ij}\varphi_j (t)},\quad i=1,\hdots,N-1\\
\varphi_j (t+1) &= \frac{\ell_j}{\sum_{i=0}^{N-1} q_{ij} \psi_i(t+1)},\quad j=1,\hdots,N-1.
\end{align}
with the initial values given by: 
\begin{equation}
\psi_i (0) = a_i,\quad \varphi_j (0) = \ell_j,\quad i,j=0,\hdots,N-1
\end{equation}
Finally, the solution of the optimization problem can be written as:
\begin{equation}
x_{ij} = q_{ij} y_{ij}= q_{ij} \psi_i \varphi_j,\quad i,j=0,\hdots,N-1.
\end{equation}
We call this algorithm the Sparse RAS (SRAS) algorithm, 
since it provides a solution to the \textit{Sparse ME} (SME) problem. 
The solution is the closest sparse matrix $x$ to the adjacency matrix $q$,  
which is also consistent with the constraints (1).

The pseudo-code of the SRAS algorithm is given in Algorithm 1. 
In the first step the initial values for $\psi_i$ and $\varphi_j$ are set to $a_i$ and respectively $\ell_j$. 
The main loop for calculating $\psi_i$ and $\varphi_j$  is executed until the Euclidean distance ($\eta$ in the pseudo-code) 
between two consecutive states $\lbrace \psi_0,\hdots,\psi_{N-1},\varphi_0,\hdots,\varphi_{N-1}\rbrace$, corresponding 
to a complete iteration, is smaller that a prescribed error $0< \delta \ll 1$: 
\begin{equation}
\eta = \left[ \sum_{i=0}^{N-1} \left[ \psi_i(t+1) - \psi_i(t) \right] ^2 + \sum_{j=0}^{N-1} \left[ \varphi_j(t+1) - \varphi_j(t) \right] ^2 \right]^{1/2} \leq \delta.
\end{equation}

\begin{algorithm}[t!]
\caption{SRAS: calculates a sparse ME solution.}
\begin{algorithmic}
\STATE $\textbf{function} \text{ SRAS}(N, a, \ell, q, \delta)$
\bindent
\STATE $x \leftarrow$ array$(N^2)$
\STATE $\psi \leftarrow$ array$(N)$
\STATE $\varphi \leftarrow$ array$(N)$
\FOR{$i=0:N-1$}
	\STATE $\psi_i \leftarrow a_i$
	\STATE $\varphi_i \leftarrow \ell_i$
\ENDFOR
\STATE $\eta \leftarrow \infty$
\WHILE{$\sqrt \eta > \delta$}
	\STATE $\eta \leftarrow 0$
	\FOR{$i=0:N-1$}
		\STATE $s \leftarrow 0$
		\FOR{$j=0:N-1$}
			\STATE $s \leftarrow s + q_{ij} \varphi_j$
		\ENDFOR
		\STATE $\xi \leftarrow a_i / s$ 
		\STATE $\eta \leftarrow \eta + (\xi - \psi_i)^2$ 
		\STATE $\psi_i \leftarrow \xi$
	\ENDFOR
	\FOR{$j=0:N-1$}
		\STATE $s \leftarrow 0$
		\FOR{$i=0:N-1$}
			\STATE $s \leftarrow s + q_{ij} \psi_i$
		\ENDFOR
		\STATE $\xi \leftarrow \ell_j / s$
		\STATE $\eta \leftarrow \eta + (\xi - \varphi_j)^2$ 
		\STATE $\varphi_j \leftarrow \xi$
	\ENDFOR
\ENDWHILE
\FOR{$i=0:N-1$}
	\FOR{$j=0:N-1$}
		\STATE $x_{ij} \leftarrow q_{ij} \psi_i \varphi_j$
	\ENDFOR
\ENDFOR
\RETURN $x$
\eindent
\STATE $\textbf{end function}$
\end{algorithmic}
\end{algorithm}

\noindent
The final solution is calculated according to the equation (32), using the adjacency matrix 
$q=[q_{ij}]$ and $\lbrace \psi_0,\hdots,\psi_{N-1},\varphi_0,\hdots,\varphi_{N-1}\rbrace$. 

We should also note that the complexity of the main loop of the SRAS algorithm is $O(2N^2)$, while the 
complexity of the main loop of the RAS algorithm is $O(4N^2)$. Thus, SRAS is twice faster than RAS, 
due to the fact that the iterated variables $\psi$ and $\varphi$ are one dimensional arrays of size $N$ (SRAS), 
comparing to $x$ which is an array of size $N^2$ (RAS).

Given the correct adjacency matrix $q$ of the interbank network, the SRAS algorithm solves the problem exactly, 
by providing the closest matrix $x$ satisfying the constraints (1), such that $q = \Theta(x)$. 
Unfortunately, in reality we do not know the correct adjacency matrix $q$, so we can only "guess" such a matrix 
by assuming that we can estimate the connectivity $\kappa$.
Therefore, the problem we are facing is to find an adjacency matrix $q$ with a given connectivity $\kappa$. 
Such an adjacency matrix will play the role of "support" for a candidate solution matrix $x$, 
even though it may not be the real adjacency matrix.

Since the constraining values (1) are strictly non-negative, $a_i>0$ and $\ell_j>0$, $i,j=0,\hdots,N-1$, the minimum connectivity must be $\kappa_{min}=1/N$. 
Also, since the elements on the main diagonal must be always equal to zero, the maximum connectivity is $\kappa_{max}=1-1/N$.
We should note also that each row and column of $q$ must contain at least a nonzero element, otherwise $q$ cannot 
be a support for $x$, which means that we must have: 
\begin{equation}
\sum_{j=0}^{N-1} q_{ij} \geq 1,\quad \sum_{i=0}^{N-1} q_{ij} \geq 1, \quad i,j=0,\dots,N-1.
\end{equation}
Of course this is only a necessary condition, and it does not guarantee that the candidate solution $x$ will satisfy the imposed constraints (1). 
However, if this condition is satisfied, then the matrix $q$ can play the role of a support for a candidate solution. 

\begin{algorithm}[t!]
\caption{ADJACENCY-MATRIX: calculates a random adjacency matrix.}
\begin{algorithmic}
\STATE $\textbf{function} \text{ ADJACENCY-MATRIX}(N, \kappa, t_{max})$
\bindent
\STATE $q \leftarrow$ array$(N^2)$
\FOR{$i=0:N-1$}
	\FOR{$j=0:N-1$}
		\STATE $q_{ij} \leftarrow 0$
	\ENDFOR
	\STATE $p_{i} \leftarrow i$
\ENDFOR
\FOR{$i=N-1:1$}
	\STATE $j \leftarrow \text{rand}(i)$ 
	\STATE $m \leftarrow p_i$
	\STATE $p_i \leftarrow p_j$
	\STATE $p_j \leftarrow m$
\ENDFOR
\FOR{$i=0:N-1$}
	\STATE $q_{ip_i} \leftarrow 1$
\ENDFOR
\FOR{$n=N:\kappa N^2-1$}
	\STATE $m \leftarrow \text{rand}(N^2)$
	\STATE $j \leftarrow \lfloor m/N \rfloor$
	\STATE $i \leftarrow m-jN$
	\WHILE{$q_{ij} = 1 \textbf{ or } i = j$}
		\STATE $m \leftarrow (m+1) \text{mod} N^2$
		\STATE $j \leftarrow \lfloor m/N \rfloor$
		\STATE $i \leftarrow m-jN$
	\ENDWHILE
	\STATE $q_{ij} \leftarrow 1$
\ENDFOR
\RETURN $q$
\eindent
\STATE $\textbf{end function}$
\end{algorithmic}
\end{algorithm}

A random adjacency matrix $q$ with a given connectivity $\kappa$ can be constructed with the Algorithm 2. 
First we notice that any permutation matrix satisfies the minimum connectivity $\kappa_{min}=1/N$ requirement, since it has 
exactly one entry of 1 in each row and each column and 0s elsewhere. Therefore we initialize $q$ with a random permutation matrix. 
We should note that the permutation matrix must have the diagonal elements equal to zero. 
Such a random permutation matrix can be obtained using a simple modification of the Fisher-Yates shuffle algorithm \cite{key-14}, 
applied to the lines (or columns) of the identity matrix. This way $q$ becomes a permutation matrix with 0s on the 
diagonal. The algorithm continues by flipping randomly $\kappa N^2-N$ 0s to 1, such that in the end there are $\kappa N^2$ 1s in the matrix. 
Only the positions off the main diagonal are used in this process. 
The pseudo-code is given in Algorithm 2, and it solves the problem exactly 
for $1/N \leq \kappa \leq 1-1/N$. 
Here $\lfloor . \rfloor$ is the floor integer function, and the $\text{rand(m)}$ function returns a uniform random integer in $\lbrace 0,...,m-1 \rbrace$.

\section{Sparse reconstruction numerical results}

In this section we discuss the performance of the SRAS algorithm when we only know the connectivity $\kappa$ of the adjacency matrix and  
the randomly generated constraint values $a_i>0$ and $\ell_j>0$, $i,j=0,\hdots,N-1$.  
In this case we first generate a support adjacency matrix $q$ with the given connectivity $\kappa$, and then we calculate the 
candidate solution $x$ using the SRAS algorithm. 

The above procedure does not guarantee that the SRAS algorithm 
will converge to the correct solution, because it is very likely that the randomly generated adjacency matrix support 
is not the correct one. 
We should note that this case is different from the case when the correct adjacency matrix is known, and SRAS can find the solution $x$ exactly. 
This is a consequence of the fact that not all the sparse adjacency matrices $q$, with a given connectivity $\kappa$, can provide support to a 
candidate matrix $x$ compatible with the constraints (1). 
Therefore, the main question is: what is the minimum connectivity $\kappa^{*}$ for the randomly generated adjacency matrices $q$ that can provide the 
correct support for a candidate solution $x$  
satisfying the linear constraints (1)?  
Therefore, we expect that there is a connectivity range $[\kappa^{*}, \kappa_{max}]$ for randomly generated adjacency matrices $q$, that can also provide a support 
for the solution matrices $x$ which are compatible with the constraints (1). 
Here, $\kappa_{max}=1-1/N$ corresponds to the dense network reconstruction case, which always admits a compatible solution, however it is not 
clear how to find $\kappa^{*}$, which corresponds to the minimum critical connectivity for which we can randomly choose a support $q$ such that the resulted 
candidate solution $x$ satisfies the constraints (1). 

In order to find $k^{*}$ numerically we measure the deviation from the constraints satisfaction as a function of the connectivity $\kappa$:
\begin{equation}
\varepsilon = \left[\frac{\sum_{i=0}^{N-1} \left( \sum_{j=0}^{N-1} x_{ij} - a_i \right)^2 + 
\sum_{j=0}^{N-1} \left( \sum_{i=0}^{N-1} x_{ij} - \ell_j \right)^2}
{\sum_{i=0}^{N-1} a_i^2 + \sum_{j=0}^{N-1} \ell_j^2}\right] ^{1/2}.
\end{equation}
We also measure the normalized entropy $S_x \in [0,1]$ of the candidate solution matrix $x$, as a function of $\kappa$:
\begin{equation}
S_x = -\frac{1}{2\ln N} \sum_{i=0}^{N-1}\sum_{j=0}^{N-1} x_{ij}\ln x_{ij}, 
\end{equation}
and the entropy $S_q \in [0,1]$ of the adjacency support matrix $q$:
\begin{equation}
S_q (\kappa) = -\kappa \log_2 \kappa - (1-\kappa) \log_2 (1 - \kappa).
\end{equation}

\begin{figure}[ht!]
\centering \includegraphics[width=15cm]{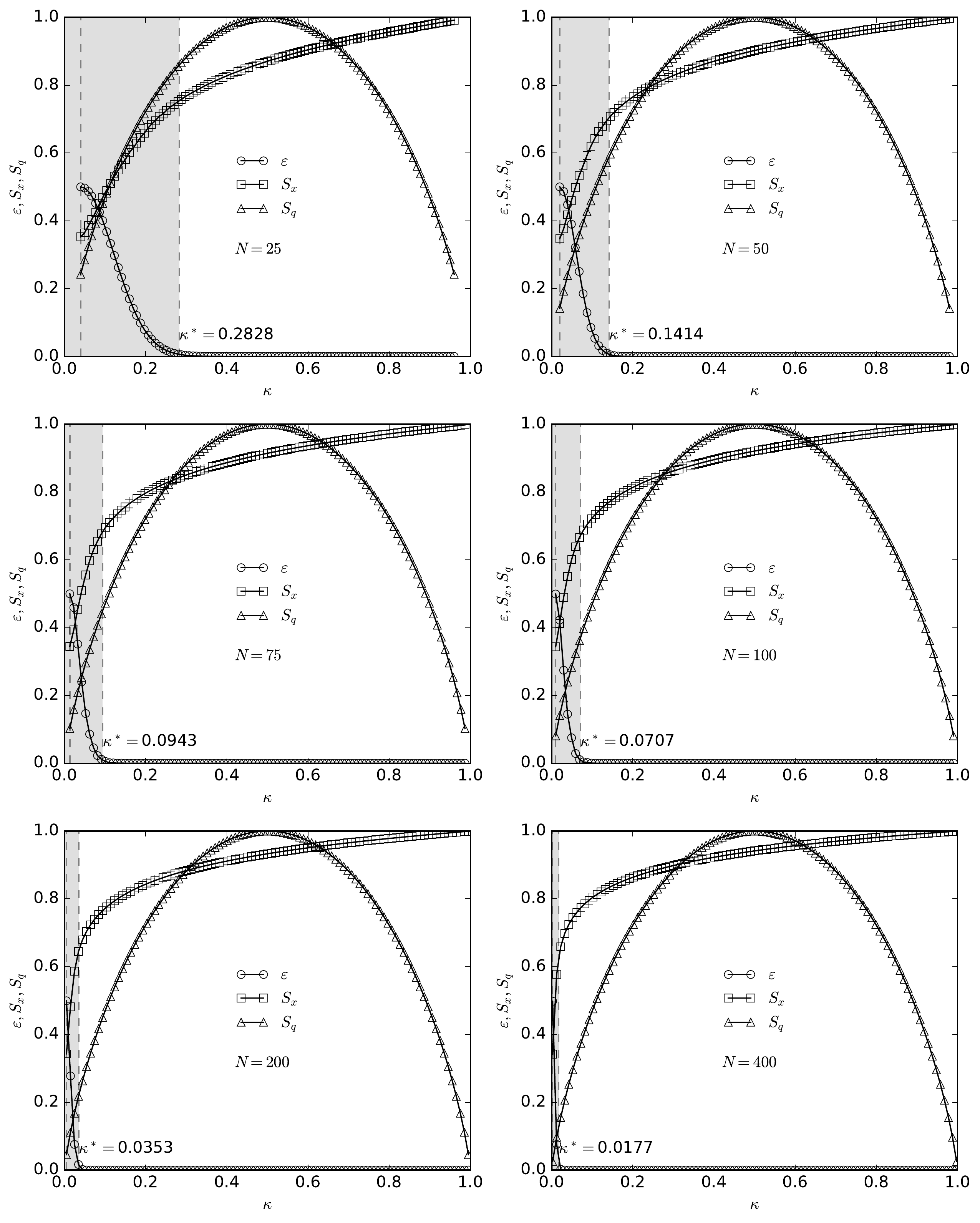}\caption{The deviation from the constraints satisfaction $\varepsilon$, 
the entropy of the solution $S_x$ and the entropy of the adjacency matrix $S_q$ for the SME method (see the text for details).}
\end{figure}

The numerical results are shown in Figure 1 for different network sizes: $N=25,50,75,100,200,400$.  
The convergence tolerance parameter for the SRAS algorithm is set to $\delta=10^{-7}$. The connectivity $\kappa$ is varied 
in the interval $[\kappa_{min}, \kappa_{max}]$, with a step $\Delta \kappa = (\kappa_{max}-\kappa_{min})/M$, 
where $k_{min}=1/N$, $k_{max}=1-1/N$ and $M=100$. Also, for each $\kappa$ we averaged the results over $10^3$ trials. 

We should note that the error $\varepsilon$ is almost exactly described by the following Gaussian function:
\begin{equation}
\varepsilon(\kappa, N) = \frac{1}{2} \exp \left[ -\frac{1}{8} \left( N \kappa -1\right)^2 \right].
\end{equation}
The continuous line on the figures is the fit with the above function. 
The gray area on the figures corresponds to the interval $[0,\kappa^{*})$ where the procedure converged to a solution $x$ inconsistent with the imposed constraints (1), 
while the white area is the interval $[\kappa^{*}, \kappa_{max}]$ where the problem was solved with an error smaller than $\varepsilon^{*} = 0.5\cdot10^{-2}$. 
The critical value $\kappa^{*}$ where this transition occurs is inverse proportional with the size of the network $N$: 
\begin{equation}
\kappa^{*}(\varepsilon^{*}, N) = \left( 1 + \sqrt{8\ln \frac{1}{2\varepsilon^{*}}} \right)N^{-1}, 
\end{equation}
and therefore the error vanishes for large networks:
\begin{equation}
\lim_{N\rightarrow\infty} \kappa^{*}(\varepsilon^{*}, N) = 0.
\end{equation}

\section{Contagion stress-test}

Once the matrix of interbank exposures $x$ is calculated, we can specify the shock that triggers the contagion. 
Usually, the contagion is simulated by letting the banks go bankrupt one at a time and measuring the number of banks that fail afterwards  
due to their direct or indirect exposure to the failing bank \cite{key-3}. Therefore, one bank failure potentially can trigger a cascade of consequent failures.

Let us suppose that $c_i(0)$ is the initial capital of the banks $b_i$, $i=0,\hdots,N-1$. 
Then the shock consists in assuming that the bank $b_i \in B$ fails 
due to some external reasons, and therefore any bank $b_j$, $j\neq i$, $j=0,\hdots, N-1$, 
loses a quantity of money equal to its exposure multiplied by a parameter $\theta \in [0,1]$ 
for loss rate (loss-given-default) \cite{key-3,key-11}. 
Therefore, if $F_t\subseteq B$ is 
the set of banks that failed at the time step $t$, then the capital of bank $j$ at time $t+1$ is:
\begin{equation}
 c_j(t+1) = c_j(t) - \theta \sum_{n \in F_t} x_{jn}.
\end{equation} 
If the loss of the bank $b_j$ exceeds its initial capital $c_j(0)$, which also means that the current capital becomes negative:
\begin{equation}
c_j(t+1) \leq 0, 
\end{equation}
then the bank $b_j$ also fails.
The contagion process stops if no additional banks fail, otherwise another round of contagion takes place.
Thus, the quantity we measure is the fraction $\xi \in [0,1]$ of the banks that failed after such a shock, 
as a function of the loss rate $\theta$ and the connectivity $\kappa$ of the interbank network:
\begin{equation}
 \xi(\theta, \kappa) = N^{-1}\vert F(\theta, \kappa) \vert.
\end{equation} 

In Figure 2 we show the simulation results for a network with $N=200$ banks. 
Here we assumed that the initial capital of the banks is $c_i=0.01$, $i=0,\hdots,N-1$, and $\Lambda=N$. The curves on the figures correspond to 
the fraction $\xi$ of the failed banks for: (i) the true exposure matrices, which are randomly generated (circles); (ii)  
the standard ME solution for the dense matrix exposure reconstruction (stars); (iii)
the SME solution for the sparse matrix exposure reconstruction (squares). 

\begin{figure}[t!]
\centering \includegraphics[width=15cm]{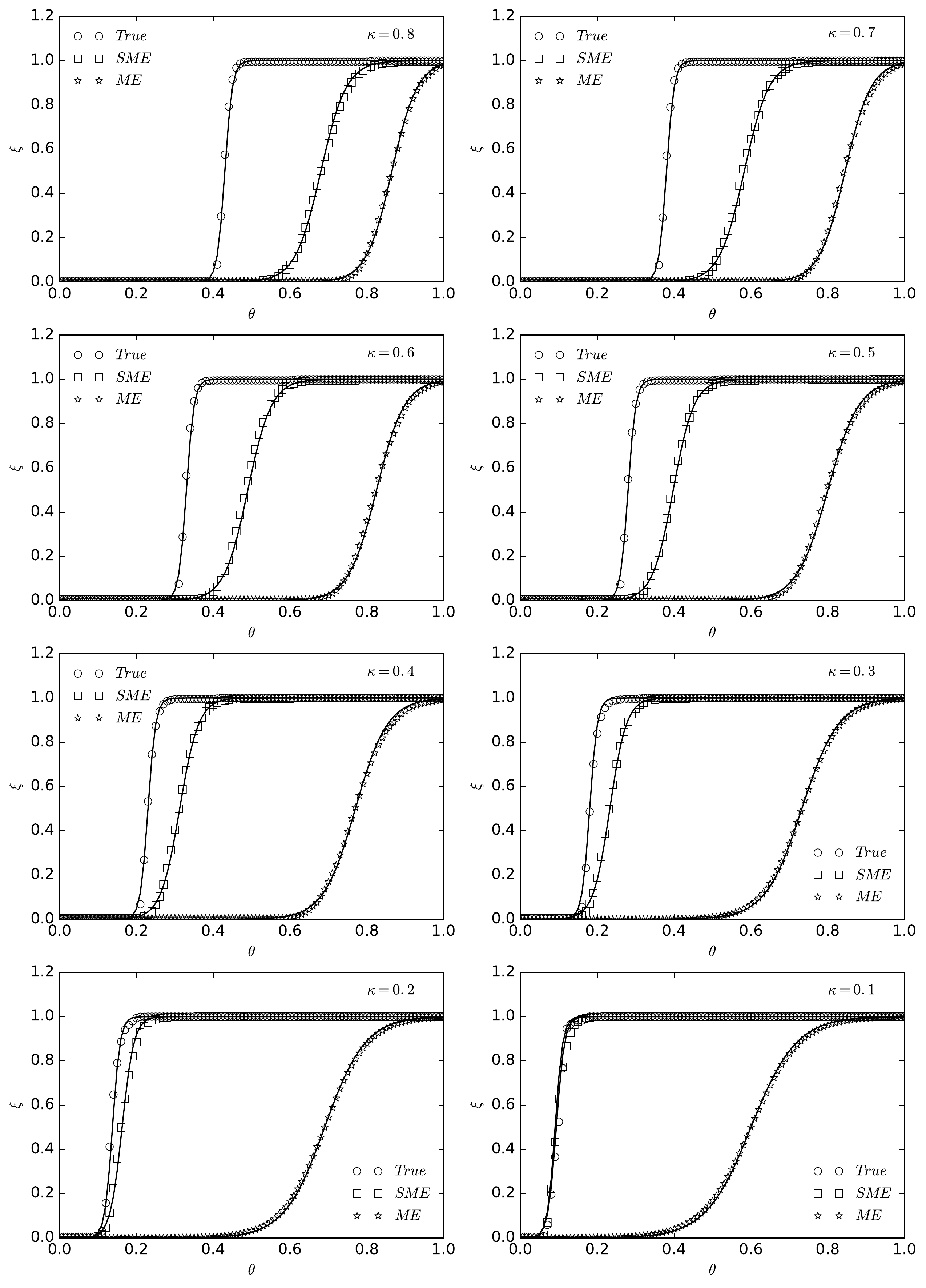}
\caption{The fraction of bank defaults $\xi$ as a function of the loss rate $\theta$ and the connectivity $\kappa$: 
the true exposure matrices (circles);  the standard ME solution (stars); and the SME solution (squares).}
\end{figure}

\begin{figure}[t!]
\centering \includegraphics[width=15cm]{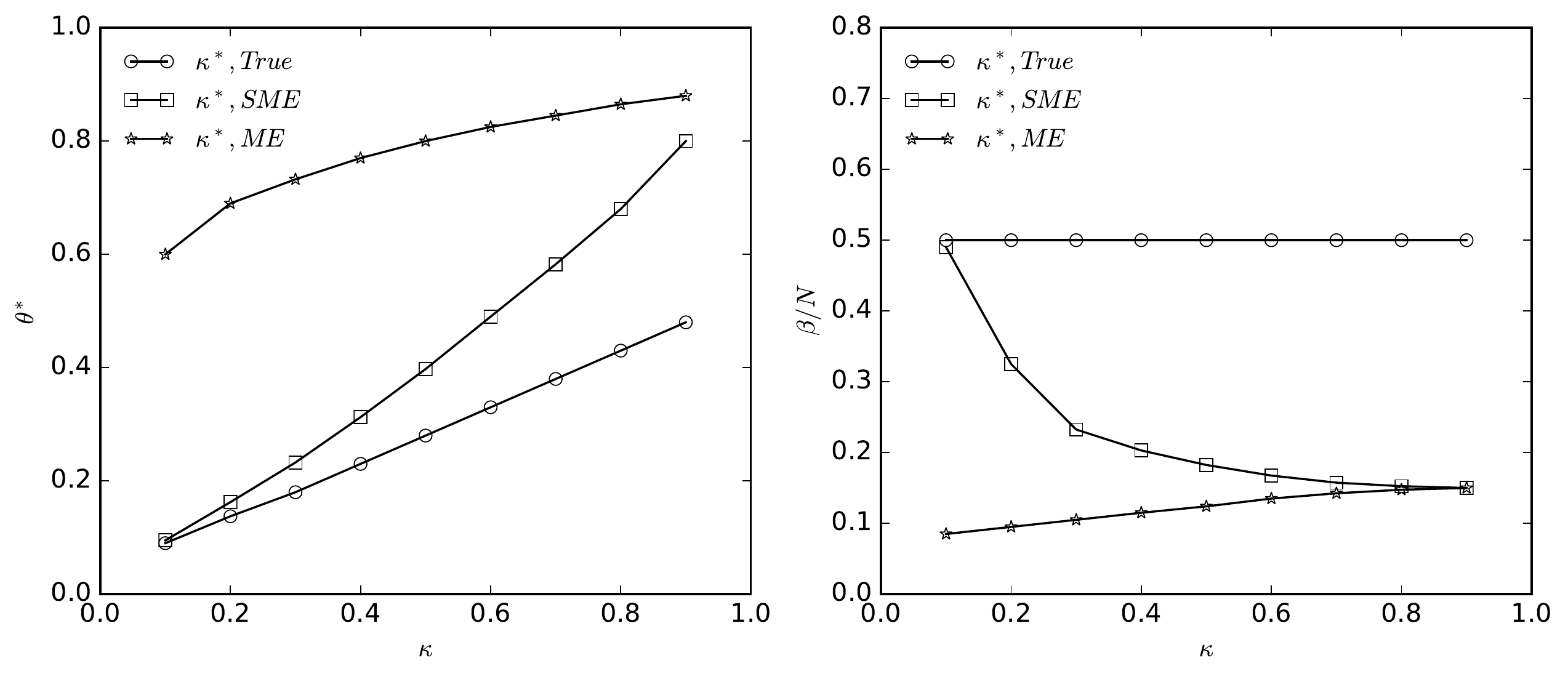}
\caption{The parameters $\theta^{*}$ and $\beta$ of the logistic growth model for the fraction of bank defaults $\xi$ 
as a function of the connectivity $\kappa$.}
\end{figure}

One can see that the fraction of defaults $\xi(\theta,\kappa)$ undergoes a phase transition as a function of $\theta$, 
which shifts towards lower values of $\theta$ when the connectivity of the network $\kappa$ decreases, as shown in Figure 2. 
In fact, the fraction of defaults $\xi$ can be described quite well by a simple logistic growth model with the solution:
\begin{equation}
\xi(\theta, \kappa) = \frac{1}{1 + \exp [-\beta (\theta - \theta^{*})]}, 
\end{equation} 
where $\theta^{*}$ is the midpoint and $\beta$ is the rate of growth (or the rate of defaults).
The logistic model solution is also shown in Figure 2, and it corresponds to the continuous lines. 
The parameters of the model are shown 
in Figure 3. One can see that for the true matrix, the midpoint has a linear dependence on $\kappa$, $\theta^{*}\simeq 0.05 + 0.5k$, 
while the rate $\beta/N = 0.5$ is constant. 
Also, one can see that the SME method provides a much more realistic result at low connectivity values than the 
standard ME method, which severely underestimates the risk of contagion. 

\section{Conclusion}

In this paper we have studied the systemic risk implied by the reconstruction of interbank exposures 
from incomplete information. We have also developed an efficient algorithm to solve the 
sparse network reconstruction problem. Also, we have shown numerically that this algorithm provides 
a more reliable estimation of the risk of contagion in the interbank network than the standard 
approach based on the ME method.
Our solution also confirms the previous results obtained by comparing the ME bilateral exposures with
those obtained on the basis of actual bilateral exposures \cite{key-9}, showing once again that the ME method 
underrates the risk of contagion. In fact, our numerical simulations show that the ME method, 
widely used in the economics literature, severely underestimates the contagion risk, while the SME method 
proposed here gives more robust results.

In a closing remark we would like to note that the simulation results of the contagion problem strongly 
depend on the network topology and on the distribution values of the bilateral exposures, and only by 
knowing the real exposures one can properly identify the actual channels for potential contagion. However, these 
simplified models and simulations, based on incomplete information, show once again that such contagion scenarios 
should not be taken too lightly, and relying on inadequate methods (such as the ME) does not provide sufficient 
warning for a potential systemic risk failure.

\end{document}